\pdfoutput=1
\documentclass[11pt,a4paper]{article}
\usepackage{array}
\usepackage[footnotesize,bf]{caption} 
\usepackage[dvips]{graphicx}
\usepackage{color} 
\usepackage{amsmath}
\usepackage{fancyhdr}
\usepackage[]{natbib}
\usepackage[LGRgreek]{mathastext}
\usepackage[yyyymmdd,hhmmss]{datetime}

\newcolumntype{L}[1]{>{\raggedright\let\newline\\\arraybackslash\hspace{0pt}}m{#1}}
\newcolumntype{C}[1]{>{\centering  \let\newline\\\arraybackslash\hspace{0pt}}m{#1}}
\newcolumntype{R}[1]{>{\raggedleft \let\newline\\\arraybackslash\hspace{0pt}}m{#1}}
\newcolumntype{T}[1]{>{\vbox to 0ex\bgroup\vfill\centering}p{#1}<{\egroup}}  

\fancyheadoffset[R]{0pt}
\fancyfootoffset[R]{0pt}

\definecolor{dred}  {rgb}{0.6,0.0,0.0}
\definecolor{aqgr}  {rgb}{0.0,1.0,0.6} 
\definecolor{viol}  {rgb}{0.8,0.6,0.8}
\definecolor{figdr} {rgb}{1.0,1.0,1.0} 
\definecolor{colnu} {rgb}{1.0,0.0,1.0} 
\definecolor{colhd} {rgb}{1.0,0.8,0.0} 

\newcolumntype{C}[1]{>{\centering\let\newline\\\arraybackslash\hspace{0pt}}m{#1}}
\newif\ifhpar
\newif\ifpbr
\pbrtrue 

\title{\bfseries{\textsc{VOCSMAT: a connectionist-inspired \\ treatment proposal
   for relational traumas}}}

\author{Alessandro Fontana} 
\date{}

\begin{document}
\maketitle
   
\clubpenalty=10000
\widowpenalty=10000

\begin{abstract}
Psychological traumas are thought to be present in a wide range of conditions, including post-traumatic stress disorder, disorganised attachment, personality disorders, dissociative identity disorder and psychosis. This work presents a new psychotherapy for psychological traumas, based on a functional model of the mind, built with elements borrowed from the fields of computer science, artificial intelligence and neural networks. The model revolves around the concept of hierarchical value and explains the emergence of dissociation and splitting in response to emotional pain. The key intuition is that traumas are caused by too strong negative emotions, which are in turn made possible by a low-value self, which is in turn determined by low-value self-associated ideas. The therapeutic method compiles a list of patient's traumas, identifies for each trauma a list of low-value self-associated ideas, and provides for each idea a list of counterexamples, to raise the self value and solve the trauma. Since the psychotherapy proposed has not been clinically tested, statements on its effectiveness are premature. However, since the conceptual basis is solid and traumas are hypothesised to be present in many psychological disorders, the potential gain may be substantial.
\end{abstract}


\section{Introduction}  

The field of ``psychotraumatology'' is concerned with the study of psychological traumas of different kinds and the associated mental structures. The initial response to trauma is represented by dissociation, which can be described as ``the escape when there is no escape'' \citep{putnam1992} and often involves an integrative failure in consciousness and memory. In the long term, the most common outcome of severe traumatic experiences is a characteristic syndrome called ``post-traumatic stress disorder'' (PTSD) \citep{andreasen2010}, which can be ``complex'' when the traumas are repeated over a long period of time \citep{herman1992}. 

However, unresolved traumas are hypothesised to be present in a broader set of psychological conditions. The ``freezing'' behaviour of children with disorganised attachment is reminiscent of a dissociative process induced by an inappropriate interaction style of the caregiver \citep{main1986}. A history of childhood traumatic experiences is thought to be a contributing factor for dissociative identity disorder \citep{schaefer2008}, personality disorders \citep{elices2015, fenske2009, lowen2004} and eating disorders \citep{phillips2004}. 

Psychological traumas of various kinds may be involved in the pathogenesis of bipolar disorder \citep{Aas2016} as well as of schizophrenia \citep{horan2003}. Unresolved childhood traumas and PTSD symptoms may also contribute to the age-related cognitive decline \citep{burri2013}, while less severe traumatic forms affect at all ages possibly every human being on planet earth (the reader might argue that, through such a broad use of the term ``trauma'', its meaning becomes too diluted and useless: to avoid this risk, a rigorous definition will be provided).

The most traditional treatment for psychological traumas is represented by psychodynamic therapy, which identifies the cause of mental disorders in intrapsychic conflicts originated in the past history of the patient, and uses verbal elaboration to solve the conflicts and improve the symptoms \citep{ahles2004}. Cognitive Behavioural Therapy (CBT) \citep{longmore2007} sees dysfunctional beliefs and behaviours as the root cause of psychopathology, and treats symptoms through the direct manipulations of distorted thoughts, dysfunctional emotions and maladaptive behaviours.

EMDR (Eye Movement Desensitisation and Reprocessing) \citep{shapiro2010} is a recent clinical method, effective in the treatment of some psychological conditions, including PTSD. The theoretical framework underlying EMDR links traumas to the presence of unprocessed and incomplete mnestic traces, mostly disconnected from the rest of memory. The EMDR procedure works by accessing and reprocessing such traces, allowing their integration with the rest of memory. This therapy is reported to produce results quickly and reliably.

This work presents a new psychotherapy for psychological traumas, based on a functional model of the mind, built with elements borrowed from the fields of computer science, artificial intelligence and neural networks. The rest of the paper is organised as follows: the model is introduced in section 2 and 3; sections 4 and 5 describe the treatment proposal; section 6 draws the conclusions and outlines future research directions.

\section{A model for the mind}

\subsection*{Features and hierarchical value}

Reality can be conceived as a set of situations, each characterised through a set of active \textbf{features}. Examples of simple perceptual features are: ``pyramidal shape'', ``vertical orientation'', ``colour green''. Examples of more abstract features are: ``blue swans'', ``baseball players'', ``being a plumber''. Feature activation and deactivation is a continuous process, driven by perceptual features fed from sensory stimuli and propagated to more abstract ones in real time. This occurs on a fast time scale, as the mind ``navigates'' through everyday life. For instance, if a person is walking on the beach, the features encoding the concepts of ``sand'' and ``sea'' may be active, while features such as ``traffic light'' and ``bus stop'' are likely to be inactive. 

The term ``feature'' is borrowed from the field of artificial neural networks \citep{cox2014} where, in the case of e.g. visual applications, simple features may represent basic visual characteristics (oriented edges), complex features represent parts of objects (eyes, nose, mouth, etc.), more complex features represent whole objects (faces), or scenes consisting of many objects. In this context, the term refers also to ideas and abstract thoughts: in practice, features are the components of mental life in its broadest sense.

We postulate that features are characterised by a property called \textbf{hierarchical value}, that can be high or low (to different degrees). Features such as ``honesty'' and ``health'', for example, are considered of high value, while ``deception'' and ``illness'' are generally perceived as features of low value. The value of many features is relative and varies for different persons, groups or social networks.

The determination and change of values happens by association: if a new feature with unknown value is presented in association with high (low) value features, it will itself assume a high (low) value. Since a feature in general takes part in many associations stored in the memory of everyone of us, its value will be determined by the combined effect of all associations. Value is a long term property, expected to change on a slow time scale. 

The fact that the value of features can be perceived as either high or low is known to all: the origin of value is less obvious. A clue can come from the observation of the behaviour of social animals, who pay much attention to the role of hierarchy: the organisation and functioning of packs of lions, hyenas, wolves and many other species is based on the shared knowledge of the rank of each pack member. In some circumstances, this knowledge can be a matter of life and death.

Human beings are no exception: in many situations, it is essential to know or decide who is superior and who is inferior. This can be useful, for example, when the employee of a company has to decide which tone to use to address a peer or a senior manager. Our hypothesis is that the mind is provided with an innate mechanism that allows to understand, in a social context, which are the group leaders and which are the ``followers''. To features associated to the first ones, the mind attributes a high value, to features associated to the second ones, a low value.

\subsection*{Emotions}

\begin{figure}[t] \begin{center} \hspace*{-0.50cm}
{\fboxrule=0.0mm\fboxsep=0mm\fbox{\includegraphics[width=18.00cm]{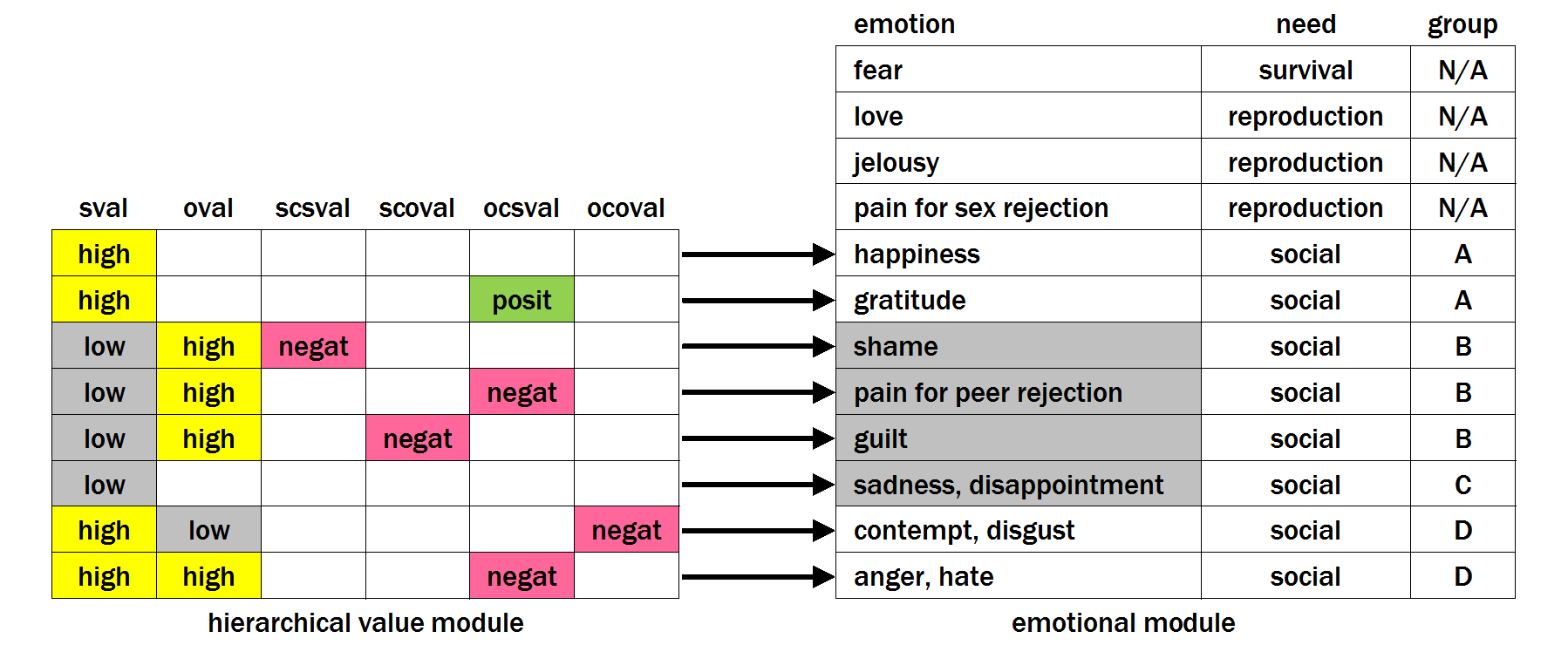}}}
\caption{Emotions, evolutionary needs and hierarchical value. Some emotions were evolved for survival and reproduction, other for social functioning. Social emotions require the computation of hierarchical value, a task performed by the hierarchical value module, which ``adds'' value to features and computes the value of self and object (sval, oval), that can be either high or low, and four other variables (scsval, scoval, ocsval, ocoval), that can be either positive or negative. Based on this computation, the emotional module produces a specific emotion (emotions associated to a low self value are shaded in grey).}
\label{emotpoles}
\end{center} \end{figure}

Many theories on the role and origin of emotions exist \citep{izard2009}, as well as computational models of the process of emotion generation \citep{marsella2010}. We hypothesise that emotions were evolved to serve three basic needs: survival (task involving one individual), reproduction (task involving two individuals) and social functioning (task involving many individuals). The first two needs are a direct translation of the requirements of Darwinian evolution and are shared by most living beings. The third derives from the necessity to organise a population of individuals and is also very common in the animal kingdom, probably because social organisation increases the Darwinian success of the species. Fig.~\ref{emotpoles} reports an (incomplete) list of the most common human emotions.

\textbf{Fear} is probably the most primitive emotion and serves the first need: avoiding physical damage and escaping death. The expression of fear does not require a relational or social context: we can have fear even if we are alone in the world. Fear (or anxiety) plays also another special role among emotions: it serves to avoid potentially harmful events (fear of dying), but also any other ``bad'' emotion: fear of shame, fear of guilt, fear of pain, etc. (this is the feeling experienced during e.g. public speaking).

\textbf{Love, jelousy and romantic pain} (the pain caused by loss of a romantic /sexual partner, by either abandonment or death) serve the second need: finding and keeping a partner for reproduction. The expression of such emotions makes sense only in a relational context, but does not require a social one: as portrayed in numerous movies, we can feel love even if we are stranded with our partner on a deserted island. 

Social emotions serve the third need: preserving the organisation of a society of individuals. These emotions are often produced as a reaction to another person's actions in a social context and depend on the assessment of who is superior and who is inferior, who is right and who is wrong. Our hypothesis is that social emotions are synthetic social /hierarchical judgements, and that the determination of hierarchical value (described previously) constitutes a precondition for the generation of such emotions.

More specifically, we foresee the existence of a module in the human mind dedicated to the computation of hierarchical value (Fig.~\ref{emotpoles}, bottom left). This module assigns value to active features and calculates \textbf{sval} (self value) and \textbf{oval} (object value), that can be either high or low, and four other variables, that can be either positive or negative: \textbf{scsval} (self contribution to self value), \textbf{scoval} (self contribution to object value), \textbf{ocsval} (object contribution to self value), \textbf{ocoval} (object contribution to object value). Self value and object value are in turn determined by the sum of the values of all \textit{active} features associated to them, $fsval_{i}$ and $foval_{i}$. In formulas:

\begin{table}[h!]
\vskip 0.25cm
\center{
\begin{tabular}{C{8cm} C{8cm}}
$sval = \sum_{i}fsval_{i}$ & $oval = \sum_{i}foval_{i}$
\end{tabular}}
\vskip 0.25cm
\end{table}

As we said, the self value depends on the set of self-associated active features. Since the set of these features changes depending on the situation, so does the self value (as well as the object value). Therefore, in our model the self value (or self-esteem) is a spatio-temporally local concept: one person can have a high self value in a given situation and a low self value in another one. We can still define a global self value as the mean self value across all situations. If the self is stably associated to a core feature set in the long term, the mean value of such features determines what we might call the person's \textbf{mood}, which can be high or low.

Based on these variables, the emotional module (Fig.~\ref{emotpoles}, bottom right) generates an emotion.
\textbf{Happiness}, for example, is generated when sval (self value) is high. \textbf{Gratitude} is generated when sval is high and ocsval is positive (in other words: when the self value is high thanks to an action done by the object). \textbf{Anger} is produced when sval is high, oval is low and ocsval is negative (in other words: when the object's action causes a decrease in self value).

\textbf{Shame} occurs when sval is low, oval is high and scsval is negative (in other words: when the self has a value lower than the object and performs an action that reduces its own value). A high object value is a precondition for the elicitation of shame: we don't feel ashamed if our mistakes are witnessed by a person who has a low hierarchical value (e.g. a small child). \textbf{Guilt} is generated when sval is low, oval is high and scoval is negative (in other words: if our actions cause a decrease in object value). \textbf{Sadness} is generated when sval (self value) is low.

So far it has been assumed that emotions depend on self and object values. We could ask if emotions depend not only on value, but also on value \emph{change}. Our intuition seems to suggest that emotions do indeed depend also on value change: disappointment seems e.g. to arise when something bad happens, causing a change (a decrease) in self value. Our answer is that the value changes that are associated to the emotion generated are in fact due to the change of situation, and emotions only depend on values (which, we recall, are spatio-temporally local).

We refer to social emotions as to ``poles'' (e.g.: pole of shame, pole of anger, etc.) and for convenience we divide them in four groups: A (emotions characterised by high self value), B (emotions characterised by low self value and high object value), C (emotions characterised by low self value) and D (emotions characterised by high self value and low object value). As we shall see, only emotions belonging to groups B and C (in which the self value if low) can be involved in a trauma. The association of shame, guilt and sadness to a lower social status is confirmed by numerous studies \citep{stevens1996}, as well as the association of anger to a higher social status \citep{tiedens2011}.

The author is aware that the proposed scheme is not complete and is not able, in its current form, to make sense of the complexity of human emotions. The main concept we wish to convey is the idea that social emotions are the outcome of a computational process involving the hierarchical value of self and others, rooted in the evolutionary need of social animals to establish a hierarchy among group members. Taking into account these limitations, we think it can still be useful to provide an intuition about the origin and purpose of emotions.   

\subsection*{Functional model of the mind, link to neurobiology}  

\begin{figure}[t] \begin{center} \hspace*{-0.00cm}
{\fboxrule=0.0mm\fboxsep=0mm\fbox{\includegraphics[width=17.00cm]{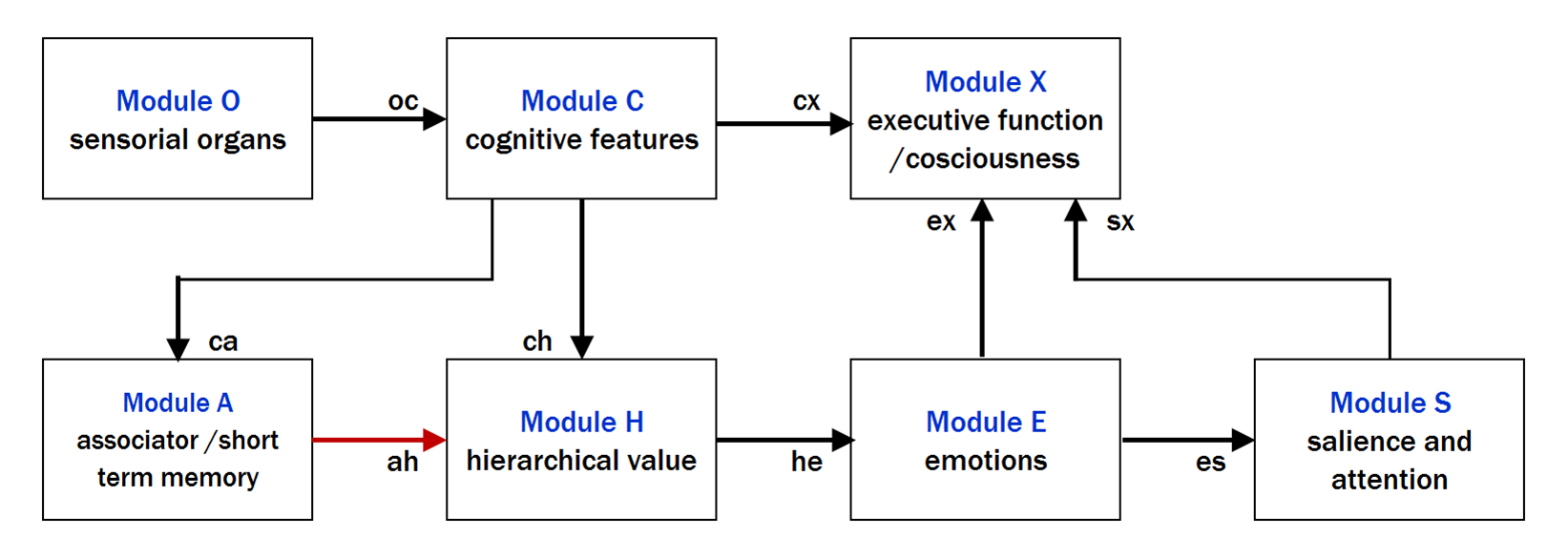}}}
\caption{Process of emotion generation. The hierarchical value module loads cognitive features with value. Based on this information, the emotional module produces an emotion that is presented to consciousness. The emotion generates salience, which is in turn presented to consciousness.}
\label{modules}
\end{center} \end{figure}

The process of emotion generation can be represented through a set of interacting modules (Fig.~\ref{modules}). Cognitive features are activated by the sensory input (link oc) and sent (ch) to the hierarchical value module, where they are loaded with value. Based on this information (he), the emotional module produces an emotion that is presented to the executive function /consciousness (ex), along with the cognitive features (cx), which are also sent in parallel to the associator (ca). Emotions influence also (es) the production of salience, which defines the importance of cognitive features and is proportional to the attention dedicated to them by the executive function (sx).

In our model of mental architecture, salience is assigned to features based on the degree of contribution to the emotion generated. If, for example, a person feels shame, because he/she has bat ears, the feature ``ears'' receives high salience. In other words, the salience module tells the mind: ``You are not far from emotion E, which mostly depends on features X, Y: these are the features that you need to monitor if you want to experience /avoid emotion E''.

Based on current neurobiological knowledge, the processing of perceptual and abstract features occurs in the cortex (in e.g. occipital cortex for visual features). The associator corresponds to the hippocampus, a structure essential for the formation of new memories. Hierarchical values could be encoded in the amygdala which, according to recent studies \citep{bzdok2013}, would encode ``good'' and ``bad'' signals and would be indispensable for understanding social hierarchy. The emotional module could be located in the striatum, while the salience module could be implemented by dopamine circuits \citep{palmiter2008, baumeister2002}. Finally, the executive function is situated in the prefrontal cortex. 

\begin{figure}[t] \begin{center} \hspace*{-0.50cm}
{\fboxrule=0.0mm\fboxsep=0mm\fbox{\includegraphics[width=18.00cm]{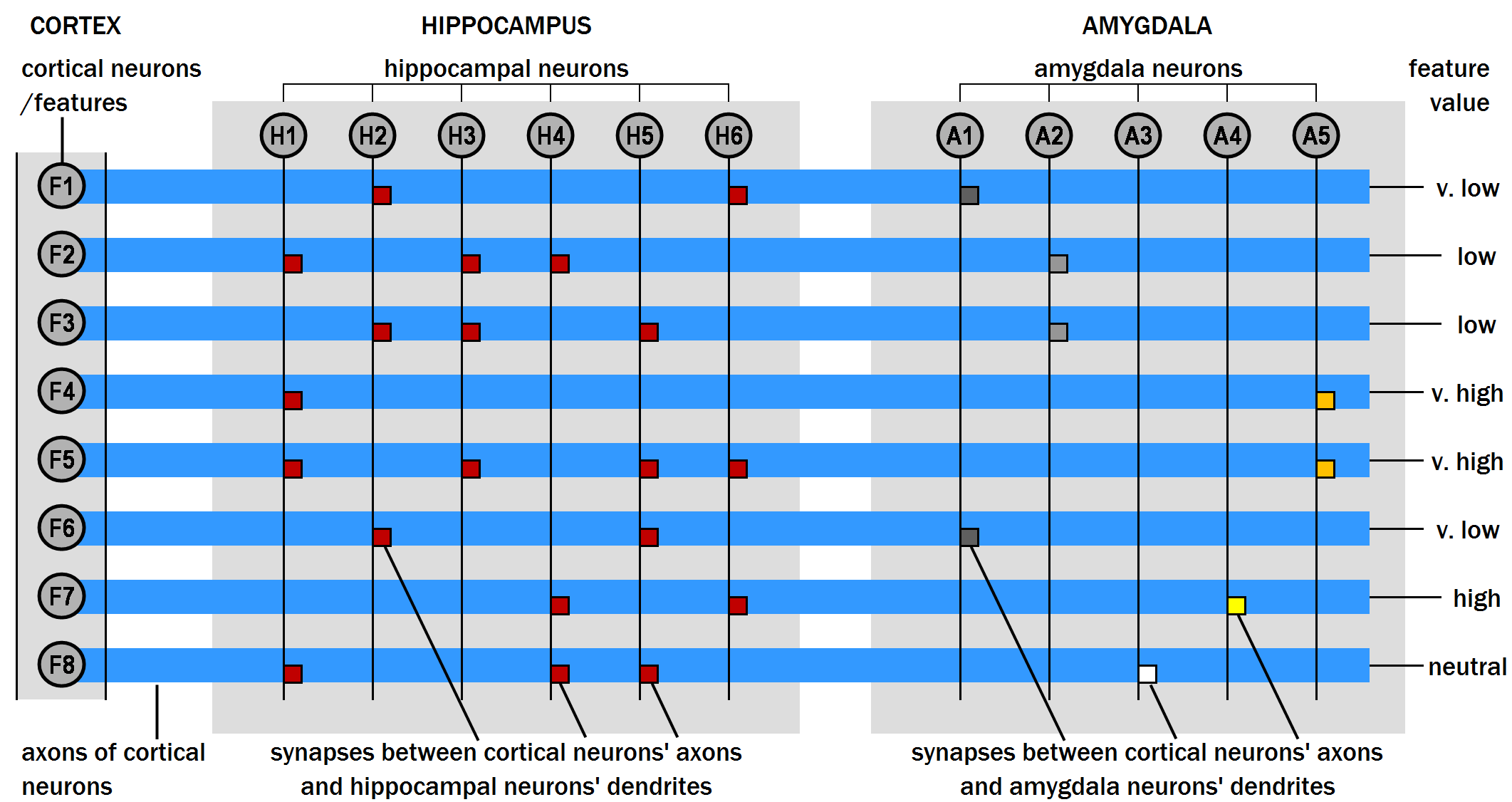}}}
\caption{Cortical neurons representing cognitive features send their axons to the hippocampus, where their co-activations are recorded in the synaptic weights between cortical neurons' axons and hippocampal neurons' dendrites. After leaving the hippocampus, the axons reach the amygdala, where the hierarchical values of features are encoded in the synaptic weights between cortical neurons' axons and amygdala neurons' dendrites.}
\label{hippamigd}
\end{center} \end{figure}

Fig.~\ref{hippamigd} shows a possible physical implementation of associator and hierarchical module in hippocampus and amygdala respectively. We hypothesise \citep{fontana2017a} that cognitive features are implemented by single neurons in the cortex (e.g. occipital cortex neurons for visual features, temporal cortex neurons for auditory features, etc.). This assumption is consistent with experiments that show how neurons exhibit selective response to very complex features, such as the face of a famous actress \citep{quiroga2005}. Cortical neurons send their axons to the hippocampus, through the entorhinal cortex acting as a ``hub'' \citep{canto2008}. After leaving the hippocampus, the path of these axons continues into the amygdala (located near the hippocampus).

In the hippocampus, the co-activations (associations) of features are recorded and stored in the synaptic weights between cortical neurons' axons and hippocampal neurons' dendrites. Each set of co-activated features constitutes a record of the ``brain dataset'', from which new, more complex features can be learned and encoded in other cortical neurons \citep{fontana2017a}. In the amygdala, the hierarchical values of features are encoded in the synaptic weights between cortical neurons' axons and amygdala neurons' dendrites. 

In our model (Fig.~\ref{modules}) the emotional module is not plastic. In other words, given the hierarchical value of active features, the emotional response is genetically determined and fixed. The plasticity of the emotional response resides in the associator and in the hierarchical module: synapses in the hippocampus are shaped by the input flow, and synapses in the amygdala are determined by the combined effect of the associations recorded in the hippocampus. If, for example, low value feature F1 co-occurs in hippocampal record H6 with high value features F5 and F7, the high value propagates from F5 and F7 to F1 (and the low value propagates from F1 to F5 and F7). As we will see later, this property represents the foundation of our therapeutic proposal.


\section{Normal and traumatic contexts}  

\subsection*{Normal functioning}  

\begin{figure*}[t] \begin{center} \hspace*{-0.50cm}
{\fboxrule=0.0mm\fboxsep=0mm\fbox{\includegraphics[width=18.00cm]{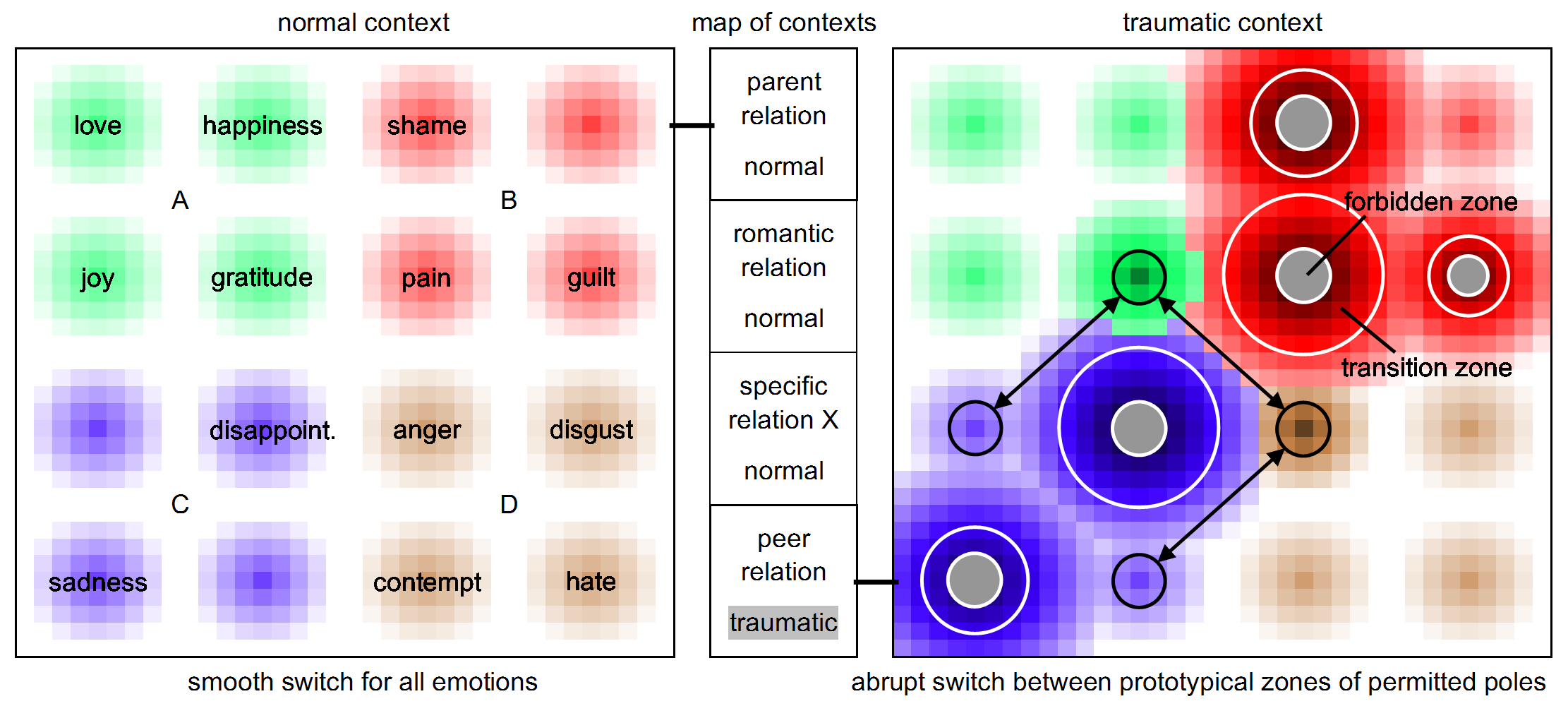}}}
\caption{Normal and traumatic contexts. The middle panel shows the map of all contexts, the left and right panels show the context plane for a normal and for a traumatic context respectively. Each point in the context plane represents a situation (defined as a set of coactive features) and is associated to an emotion (whose intensity is represented by the colour shade). In a normal context (left panel), the mind can switch smoothly between emotional poles, no pole requires dissociation and the repertoire of emotions is fully accessible. In a traumatic context (right panel), poles belonging to group B and/or group C are characterised by too intense emotional levels and are inhibited. When the mind happens to be in one of these poles (forbidden zone), dissociation intervenes. To avoid dissociation, the mind oscillates between prototypical zones of permitted poles, staying in each pole as long as the situation remains prototypical. When the situation nears the transition zone of a traumatic pole, splitting symptoms appear and the mind switches to the prototypical zone of another permitted pole.}
\label{switchx}
\end{center} \end{figure*}

Let us define a \textbf{situation} as a set of features simultaneously active. Examples of situations are: ``piano lesson with uncle'' (coactive features: image of piano, image of hands on the keyboard, sound of uncle's voice, sound of piano, etc.); ``tennis match with a friend'' (coactive features: image of racquet, image of opponent, sound of racquet hitting the ball, odour of sweat, etc.). Situations can be thought of as belonging to different \textbf{contexts}, such as ``parent relation'', ``romantic relation'', ``relation with schoolmates'', etc. (Fig.~\ref{switchx}, middle panel).

A context can be conveniently represented on a plane, where each point corresponds to an individual situation (Fig.~\ref{switchx}, left panel) and belongs to the ``zone of influence'' of an emotional pole, characterised by a specific emotion. Each pole originates from a point representing the most prototypical situation associated to the pole, and extends towards less prototypical situations. The epicentre of the anger pole, for instance, may correspond to a situation-point characterised by an object behaving very dishonestly, eliciting a very strong anger, while points further away may be characterised by a better behaviour of the object. We define a context ``normal'' if the highest emotional levels are not \emph{too} high. In this condition the mind can switch smoothly between all poles, experiencing different levels of the emotions associated to each pole (Fig.~\ref{switchx}, left panel).

\subsection*{Traumatic functioning: dissociation}

In our definition, a trauma occurs when the intensity of the elicited emotion is too high and exceeds the individual's tolerance threshold. We assume that the emotional poles that can be involved in a trauma are those in which the self value is low: shame, pain, guilt, disappointment, sadness. Emotional poles with high self value, such as love and happiness, but even anger and disgust, cannot become traumatic.  

Based on the equations reported in section 2, the production of an emotion presupposes a low self value or the existence of a value gap between object and self (object worthier than self). This in turn requires that self-associated features have on average a low value, lower than object-associated features. If a person with bat ears, big nose and thin lips considers these features of very low value, the emotions elicited when they become active are too intense and can become traumatic.

The standard response to trauma is represented by dissociation, a very complex and elusive phenomenon, investigated since the times of Pierre Janet \citep{janet1889}. Some authors \citep{nijenhuis2011} distinguish negative and positive dissociative symptoms. Negative symptoms involve functional losses such as absence of mind, depersonalisation (feeling of separation from one's body), derealisation (feeling of being detached from the world), selective amnesia and emotional detachment \citep{lanius2015, radovic2002depers}. Positive dissociative symptoms are characterised by intrusions, flashbacks, voice hearing and other manifestations that are borderline with psychosis \citep{moskowitz2008}.

At its core, dissociation is caused by an integrative failure of the different components of consciousness and memory. In the most severe cases, it can give rise to mental subsystems that behave as distinct personalities, with their own sense of self and first-person perspective. In all cases, dissociation can be considered as a distortion and an alteration of the processes of reality perception and memory formation, that protects the mind from too intense (bad) emotional levels, exceeding the coping ability of the individual. 

\subsection*{Traumatic functioning: splitting}

The space around a traumatic pole can be divided into three zones (Fig.~\ref{switchx}, right panel): the ``forbidden zone'', an area around the centre which can only be accessed in a fully dissociated state; the ``transition zone'', a safety belt around the forbidden zone; the ``free zone'', an area sufficiently far from the centre. The adoption of dissociation makes it possible for the mind the stay around a traumatic pole, keeping intolerable thoughts and emotions disconnected from consciousness and from memory. However, the disconnection of aspects of reality may hide potential dangers and have unpleasant consequences for the individual. Therefore, the mind tries to avoid traumatic poles and head towards non-traumatic poles, where the perception of reality is not restricted.

Let us assume that, in a traumatic context, the mind is initially near the gratitude pole. The mind will stay around this pole as long as conditions are prototypical, i.e. as long as the object relation is perfect, full of trust, mutual respect, etc. As the situation departs from the prototypical scenario of the gratitude pole and drifts into the transition zone of the disappointment pole, the mind switches abruptly to the prototypical zone of the anger pole. When the situation deviates from the prototypical scenario of the anger pole, the mind returns to the gratitude pole or goes to another permitted pole, and the cycle repeats itself. This corresponds to the defence mechanism of \textbf{splitting}, defined as the inability to integrate positive and negative aspects of self and others, which results is a view of the world in ``black and white'' \citep{perry2013}.

Our interpretation of splitting is broader than the traditional psychodynamic conception, in which only two ``poles'' (good and bad) are foreseen. In our vision, splitting can involve many poles (all those shown in Fig.~\ref{switchx}): the distinguishing mark of splitting is the sudden, abrupt nature of the switching between poles, which normally occurs smoothly. This is due to the presence of forbidden and transition zones around a traumatic pole that, once encountered, force the mind to jump to another pole.

We can hypothesise that, upon the first occurrence of a trauma, dissociation is total and involves all mental functions (like the ``freezing'' of disorganised attached children). Total dissociation provides a shield to pain, but may expose the individual to serious consequences in a potentially dangerous environment: it is not unrealistic to assume that, in case of repeated traumas, the mind will learn to use a selective form of dissociation which, near a traumatic situation, excludes from consciousness and memory \textit{some} emotional or cognitive channels.

This is a common phenomenon: it happens to the medical staff of intensive care units, to the employees of a slaughterhouse, to all of us. We can think of this process as of a form of learning, in which the mind selects the smallest subset of reality that needs to be dissociated to avoid emotional pain without losing touch with the ``here and now''. A malfunction of selective dissociation has been linked to schizophrenia \citep{fontana2017schizo}.      

\section{Therapeutic proposal}

A common line of thought maintains that, to resolve a trauma, it is necessary to mentally relive the traumatic event and reexperience the emotions associated, providing at the same time a more favourable interpretation of the traumatic event (e.g., ``it was not your fault'', etc.). This can be achieved with many techniques, including psychodynamic therapy, CBT and EMDR, which facilitates the reprocessing of traumatic memories. However, if the associated pain (or shame, guilt, etc.) is too strong, the patient cannot relive it (resorts to emotional dissociation) and the trauma is not resolved. 

\begin{figure}[t] \begin{center} \hspace*{-0.00cm}
{\fboxrule=0.0mm\fboxsep=0mm\fbox{\includegraphics[width=15.00cm]{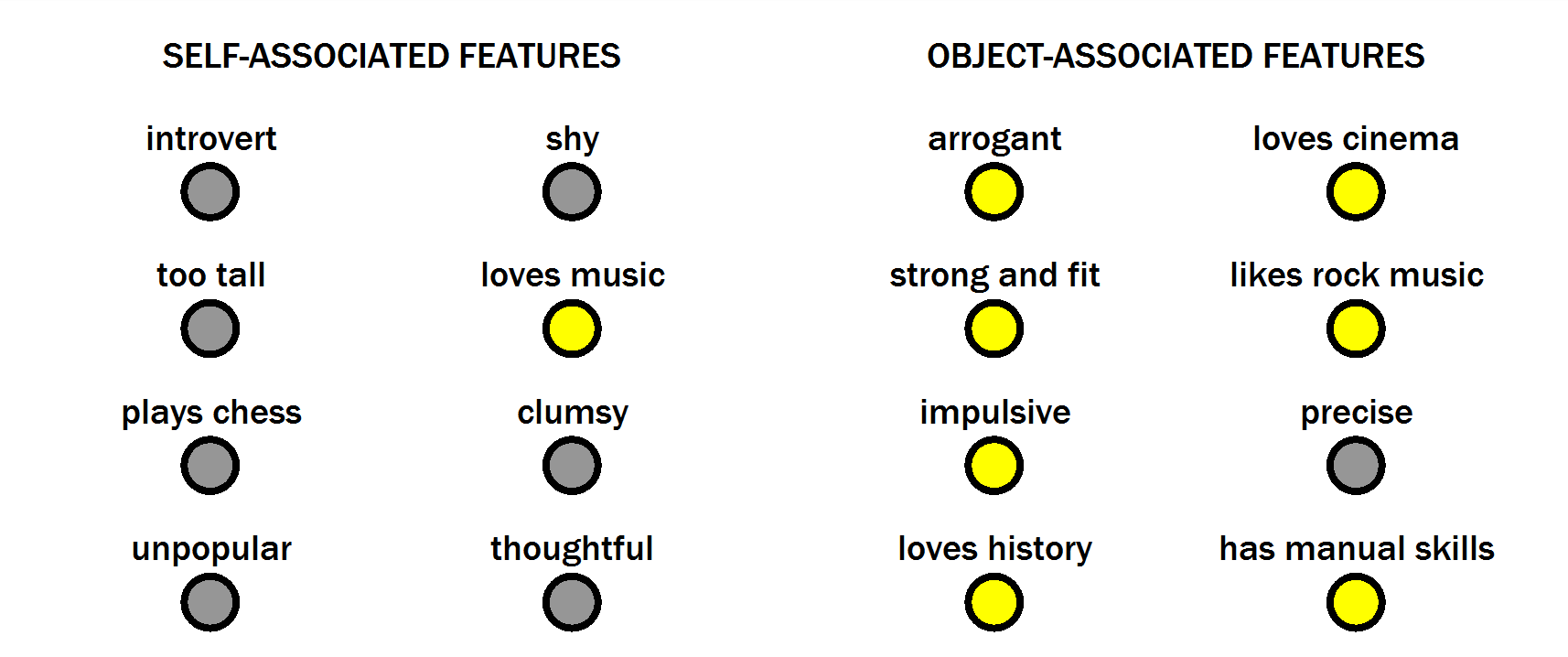}}}
\caption{On the left: self-associated features. In this example, most self-associated features are of low value. As a result, in most situations, the self has a low value. On the right: most object-associated features are of high value. As a result, in most situations, the object has a high value. This pattern is a predisposition and a precondition for the occurrence of a trauma.}
\label{featset}
\end{center} \end{figure}

Our proposal to overcome this obstacle is the \textit{preventive} reduction of the pain level. As pointed out in section 3, the intensity of emotional pain is proportional to the value difference between object and self, which in turn depends on all associated features. If the self value is low, it is because most self-associated features are of low value (Fig.~\ref{featset}, left panel); if the object value is high, it is because most object-associated features are of high value (Fig.~\ref{featset}, right panel). It is impossible to have a low-value self linked to mostly high-value features, or a high-value object linked to mostly low-value features. 

The sets of features associated to self and object represent the foundation of the traumatic structure. If the values of such features are modified, the values of self and object are expected to change accordingly and, as the value gap narrows, the level of emotional pain is expected to diminish. At this point, the trauma would be susceptible of being attacked with techniques such as EMDR, until its full resolution. Therefore, our therapeutic strategy is to target self- and object-associated features and change their values in the patient's mind.

The good news is that value is attributed to most features in a completely arbitrary way. This can be appreciated by considering the diversity of ideas, beliefs and reference values across ethnic groups, cultures, countries and historical periods. A high diversity in values can be observed also in the same historical period and in the same country, across different social networks. In a given family, the feature ``being an artist'' may be considered ``cool'' and appreciated, while ``being an engineer'' may be considered boring and worthless. For a different family, the opposite may be true. The value background of a person is initially set by the senior family members (usually the parents), but in principle nothing prevents to bring changes to it.   

This can be achieved through \textbf{counterexamples}. Assuming, for instance, that the feature ``taking risks'' has a low value, we can provide counterexamples in which a risk-taking behaviour gave good results. We might mention Julius Caesar, who chose to cross the Rubicon; Butch Cassidy and Sundance Kid, who jumped into a waterfall and saved their lives (at least in the movie!). Given the low-value feature ``failing'', we might say that Henry Ford was bankrupt a number of times before succeeding, and so on. Linking high-value examples to low-value features raises the features' value. Likewise, the association of low-value examples to high-value features can be used to lower the features' value. We call this therapeutic scheme \textbf{Value-Oriented Counterexample-Supported Massively Associative Training (VOCSMAT)}.

From the neurobiological point of view, the hypothesis is that the functioning of the emotional module (Fig.~\ref{modules}) is genetically determined and not plastic. In other words, given the features' values, the type and intensity of the emotion generated are completely determined. The plasticity lies instead in the associator and in the hierarchical module: the features' values can be modified by inserting new associations in the hippocampus, which in turn affect the coding of values in the amygdala (Fig.~\ref{hippamigd}).

Since ``negativity'' propagates by association (and is very contagious!), there might be many (several hundreds) features linked to traumas in the patient's mind: such features need to be \textit{all} addressed. For each feature, usually 3-4 counterexamples may be needed, which brings the total number of counterexamples around the figure of 1000. The proposal to use a large number of examples draws inspiration from the functioning of artificial neural networks, that learn through exposition to a huge number of examples (the so-called "big data'').

As a result of the therapy, the structure of the feature space should change from the pattern of Fig.~\ref{switchx}, right, to the pattern of Fig.~\ref{switchx}, left, turning a traumatic context into a normal one. This means that the forbidden zone and the transition zone disappear from all emotional poles, together with the need to use dissociation. Once the trauma is solved, the removal of dissociation could happen automatically, or involve a re-learning process (the mind has to learn that the once traumatic pole is now safe and dissociation is no longer needed).


The effect of dissociation can be likened to the action of weirs that separate the healthy parts of the mind in water-tight compartments. When a ``bad'' compartment is excluded, the rest of the mind /memory is made immune to the flooding of low-value features and bad emotions. Once the feature space is restructured through therapy, bad emotions become less bad and more tolerable and the weir reopens: when this happens, we can expect a sudden decrease of self value and a flood of bad emotions. However, the effect is temporary and marks the start of the integration process.
  
Framed in these terms, our therapeutic proposal might look deceptively simple. A difficulty we can expect to encounter, due to the large number of features involved, can be clarified through an example. Let us suppose to have 100 black socks that we want to turn into white socks: to achieve this goal, we apply a whitening solution to black socks, treating a given number of socks at a time. 

Let us also suppose that, at regular intervals, a random subset of socks is put into a washing machine. If the white socks in the washing machine are much fewer than the black socks (as we can expect at the beginning of the treatment), they will absorb the colour released from the black socks and turn black again. Only when the white socks become the majority, is the ``washing machine effect'' expected to be beneficial, but this may take a long time. 
 
In this metaphor, black socks correspond to low-value features, white socks correspond to high-value features and the whitening solution corresponds to the VOCSMAT treatment described above. The washing machine corresponds to the associative mechanism of the mind, that continues to function during the therapy. If the therapy could be administered ``offline'', there would be no washing machine, but this is not possible. The therapy has to be carried out ``online'', with the mind fully connected to the external world, which might remind the patient how unworthy he/she is.

The VOCSMAT therapy has some points of contacts with CBT: the key difference between the two can be illustrated with an example. Let us assume that the patient is a teenager boy, with hostile parents and no friends, convinced that these features are of low value and that he is a loser. CBT could try to convince the patient, with logical arguments, that he possesses other qualities that the others can appreciate, that there are people who love him, etc. In reality, the patient may be right in thinking that most people think he is a loser: in this case CBT tries to prove something untrue and is doomed to failure.

The VOCSMAT therapy does not try to enhance and put in evidence the high-value features of the patient (it is not needed). It accepts reality as it is and focusses on the low-value features, trying to raise their value. If the patient says: ``I'm a loser, my mother is angry at me, I don't have friends at school, everybody makes fun of me and neighbours spy on me'', the VOCSMAT therapist may say: ``Ok. So what? Even Albert Einstein was a loser at school, even Napoleon had no friends, even George Washington had a mother who ... etc. You have mental problems? You cut yourself and have other weird behaviours? You are not alone, many artists and scientists had mental problems: Charles Bukowski, John Nash, Vincent Van Gogh ...'' and so on. In this way the patient begins to value himself for what he really is in the present, and not for what he might become in the future.

Without treatment, the amount of unsolved traumas is expected to grow steadily during the course of life. Therefore, with age progression, we can expect to have an increasing number of life contexts and situations where dissociation is used and, as a result, the associative power of the mind is reduced. This would cause symptoms such as lack of attention, poor concentration, emotional detachment, anhedonia, etc., which may add to the overall age-associated cognitive decline \citep{lindenberger2014}. The proposed therapy could have an impact also on this phenomenon.  

\section{Examples of features linked to traumas and counterexamples}

This section reports lists of features related to ten (common) traumas. For four of these traumas (``making mistakes'', ``being weak'', ``being different'', ``being a loser''), counterexamples are also provided. Features can be divided into three main categories: 1) features that do not depend on the self and that can occur at any time (present and future), e.g.: ``being criticised''; 2) features that depend on the current self, e.g. ``not taking initiatives for fear of criticism'' (this category also includes pathological symptoms, e.g. risky behaviour, substance abuse, etc.); 3) features associated to an ideal self (in the future), for example: ``undertaking risky initiatives, daring''. As seen from the example, it is possible to have a feature in the second category and its opposite in the third: the value of both features must be raised (the rules of arithmetic do not apply in this case!).

Interestingly, many lists are present on the internet, that link low-value features to famous (hence high-value) persons, e.g.: ``Celebrities who were bullied before becoming famous'', ``Gorgeous stars who were dumped in the worst ways possible'', ``Famous people with disabilities'', etc. Internet users seem to be fond of such lists because, we think, these associations raise the value of self-associated low-value features in the users' minds. As a consequence, the self value improves and these persons feel better.

\ifpbr 
\clearpage \pagebreak[4] \fi
 
\begin{figure}[p] \begin{center} \hspace*{-0.25cm}
{\fboxrule=0.0mm\fboxsep=0mm\fbox{\includegraphics[width=17.50cm]{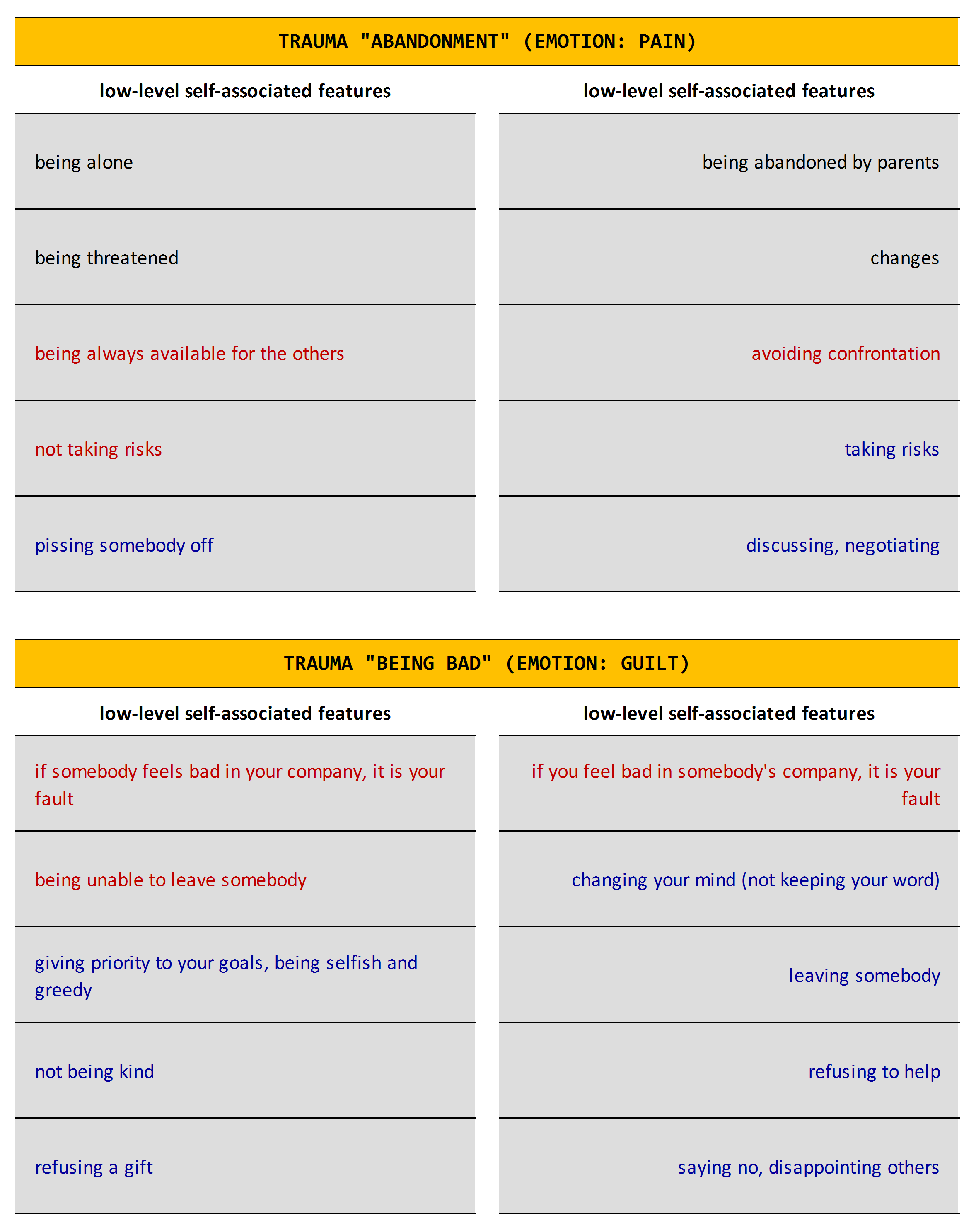}}}
\caption{Some features related to traumas ``ABANDONMENT'' and ``BEING BAD''. Features coloured in black belong to the first category (independent from self), features coloured in red to the second (linked to current self), features coloured in blue to the third (linked to future self).}
\label{abandon}
\end{center} \end{figure}

\begin{figure}[p] \begin{center} \hspace*{-0.25cm}
{\fboxrule=0.0mm\fboxsep=0mm\fbox{\includegraphics[width=17.50cm]{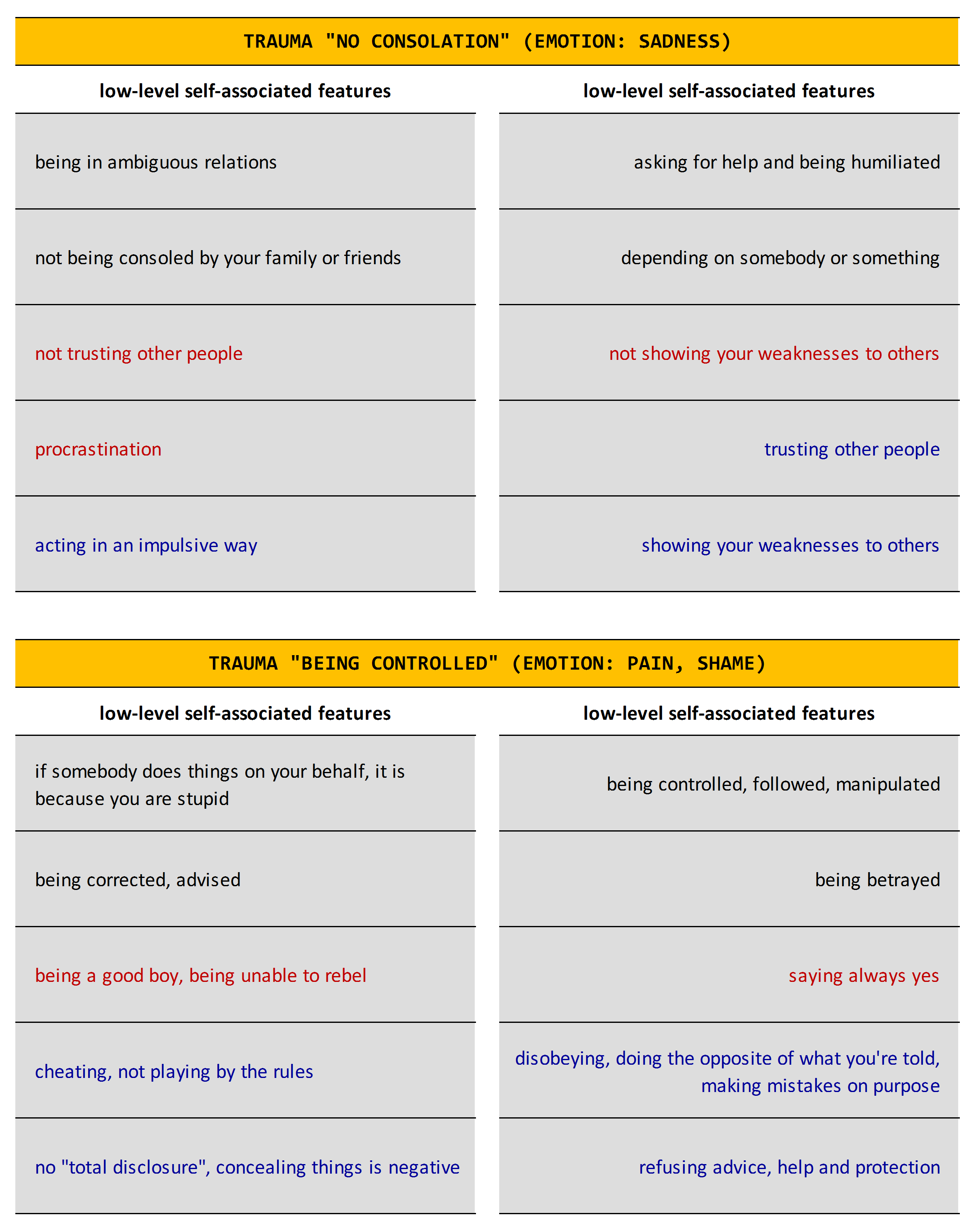}}}
\caption{Some features related to traumas ``NO CONSOLATION'' and ``BEING CONTROLLED''. Features coloured in black belong to the first category (independent from self), features coloured in red to the second (linked to current self), features coloured in blue to the third (linked to future self).}
\label{nocons}
\end{center} \end{figure}

\begin{figure}[p] \begin{center} \hspace*{-0.25cm}
{\fboxrule=0.0mm\fboxsep=0mm\fbox{\includegraphics[width=17.50cm]{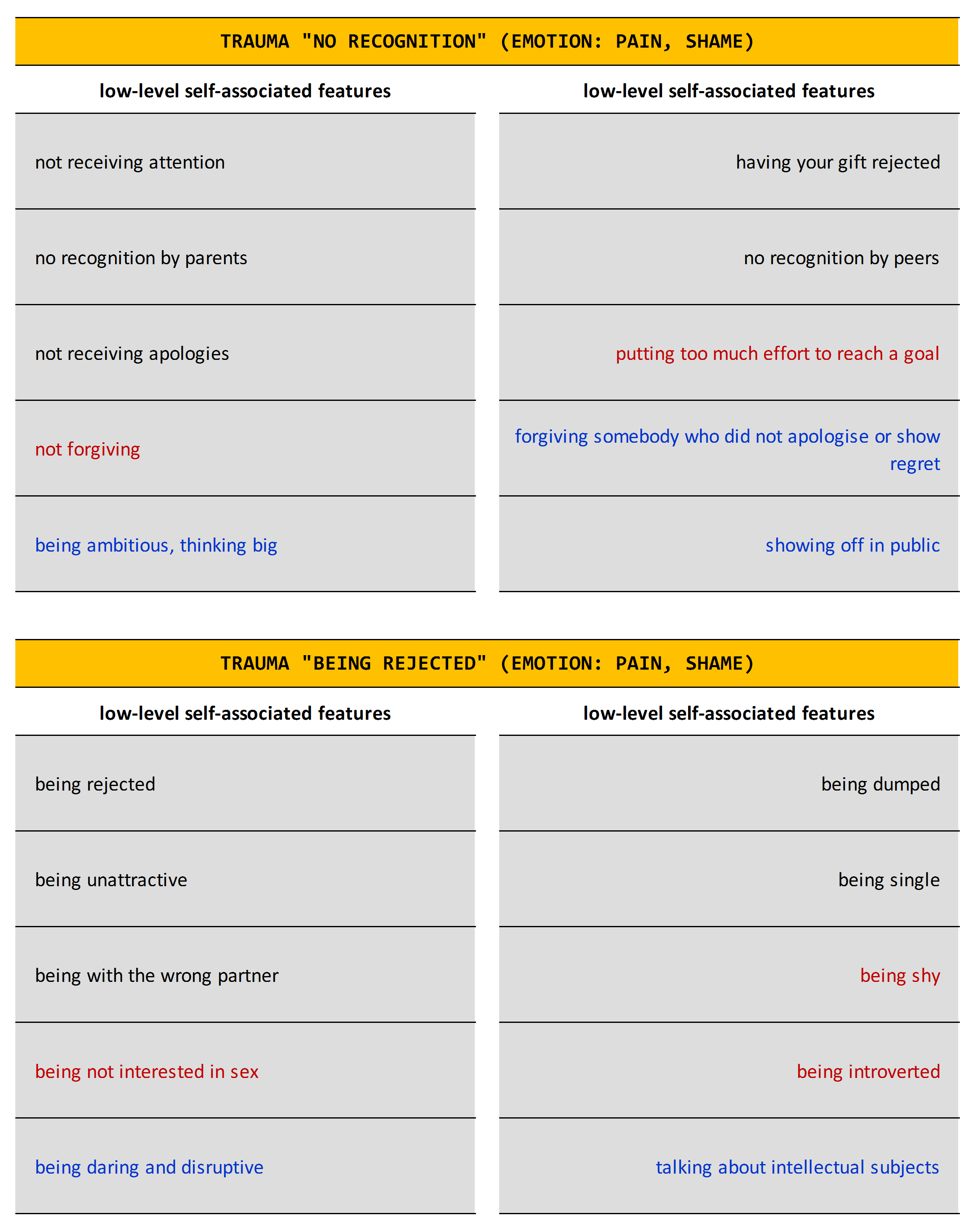}}}
\caption{Some features related to traumas ``NO RECOGNITION'' and ``BEING REJECTED''. Features coloured in black belong to the first category (independent from self), features coloured in red to the second (linked to current self), features coloured in blue to the third (linked to future self).}
\label{norecogn}
\end{center} \end{figure}

\begin{figure}[p] \begin{center} \hspace*{-0.25cm}
{\fboxrule=0.0mm\fboxsep=0mm\fbox{\includegraphics[width=17.50cm]{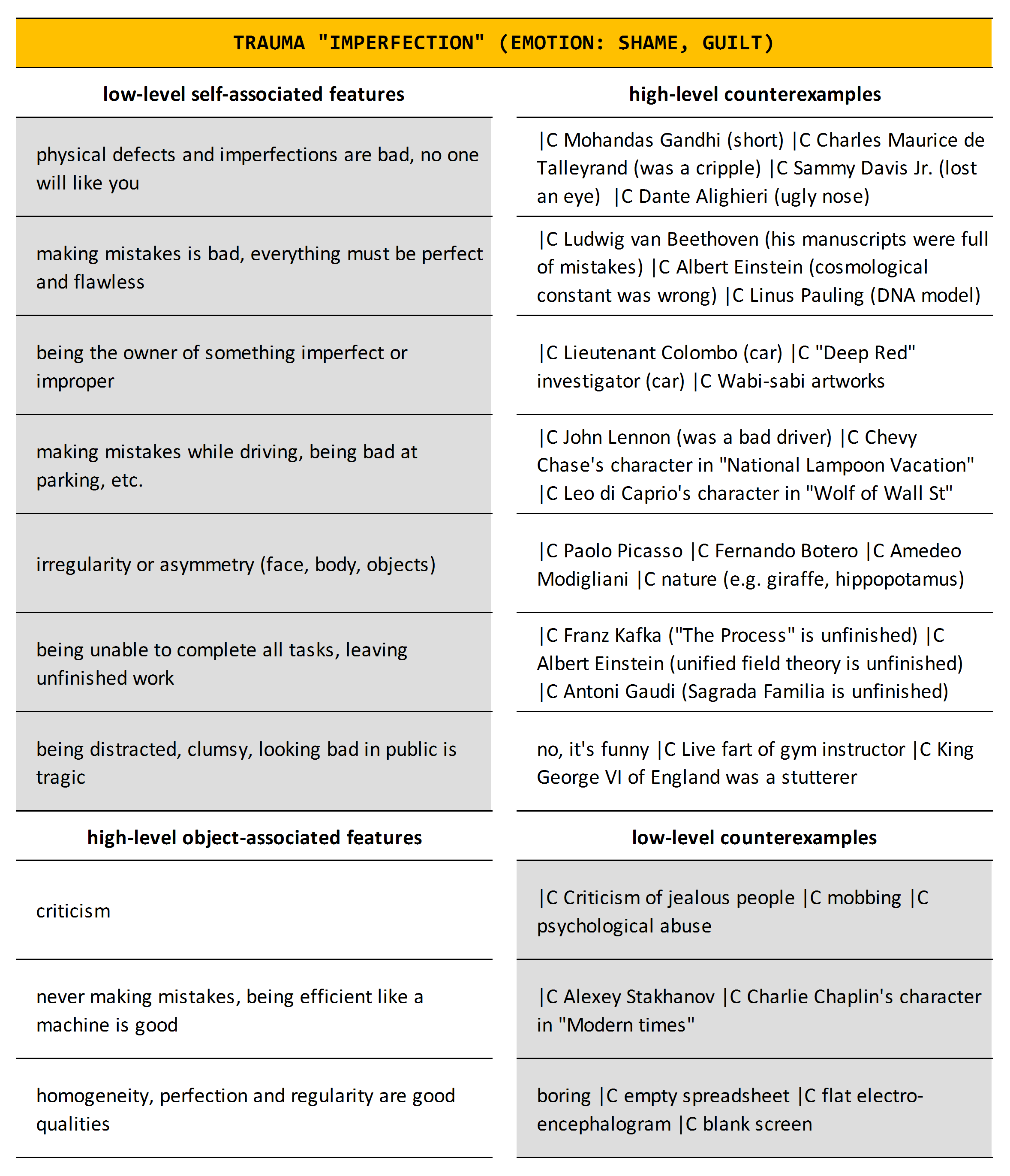}}}
\caption{Counterexamples (symbol ``$\mid$C'') for features linked to trauma ``IMPERFECTION''. The information reported is correct to the best of the authour's knowledge, does not refer to any specific ``self'' or ``object'', is intended for scientific use only and does not intend to offend any person or entity. As a further precaution, only historical figures and fictional characters are used as examples.}
\label{imperfect}
\end{center} \end{figure}

\begin{figure}[p] \begin{center} \hspace*{-0.25cm}
{\fboxrule=0.0mm\fboxsep=0mm\fbox{\includegraphics[width=17.50cm]{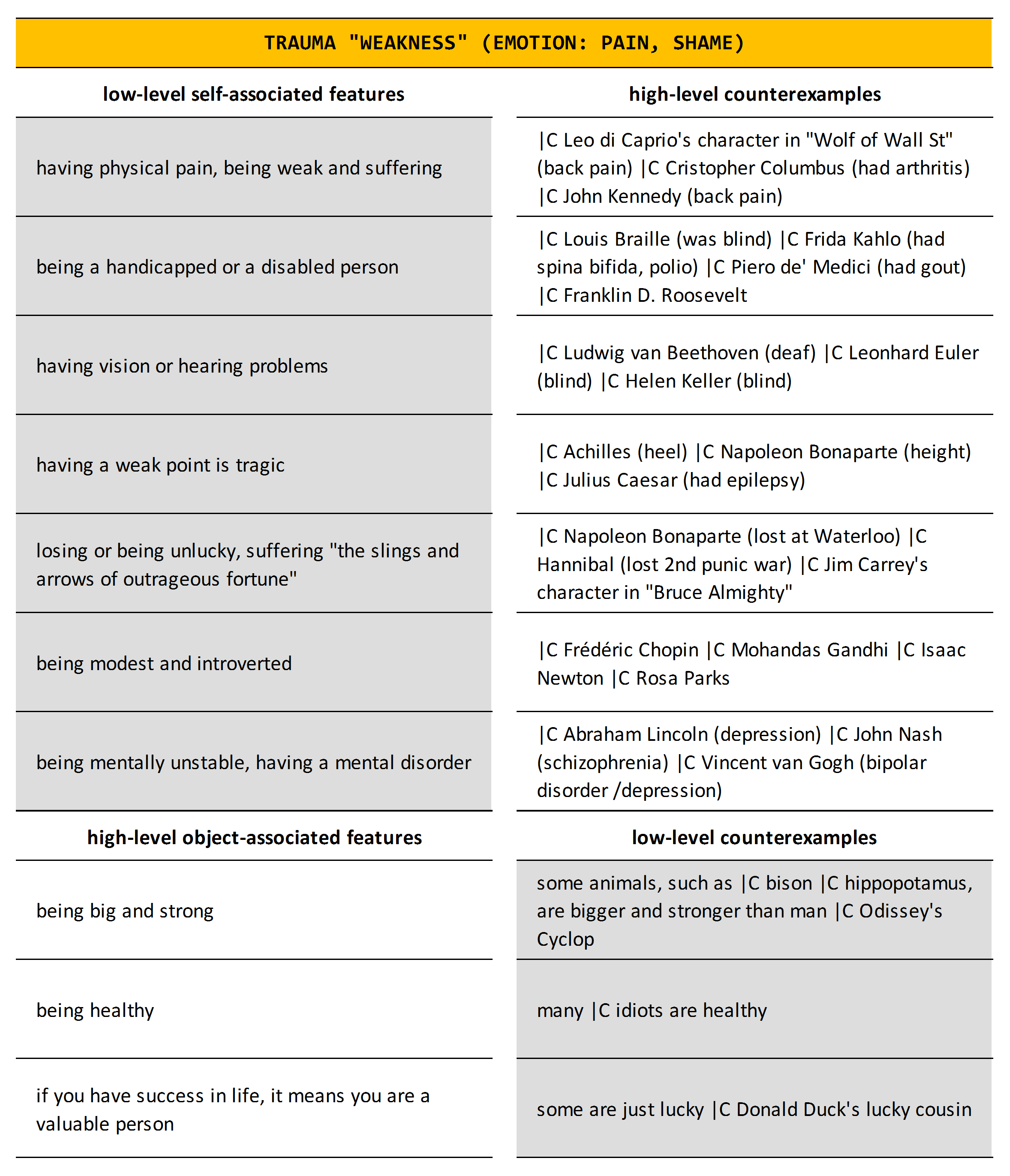}}}
\caption{Counterexamples (symbol ``$\mid$C'') for features linked to trauma ``WEAKNESS''. The information reported is correct to the best of the authour's knowledge, does not refer to any specific ``self'' or ``object'', is intended for scientific use only and does not intend to offend any person or entity. As a further precaution, only historical figures and fictional characters are used as examples.}
\label{weakness}
\end{center} \end{figure}

\begin{figure}[p] \begin{center} \hspace*{-0.25cm}
{\fboxrule=0.0mm\fboxsep=0mm\fbox{\includegraphics[width=17.50cm]{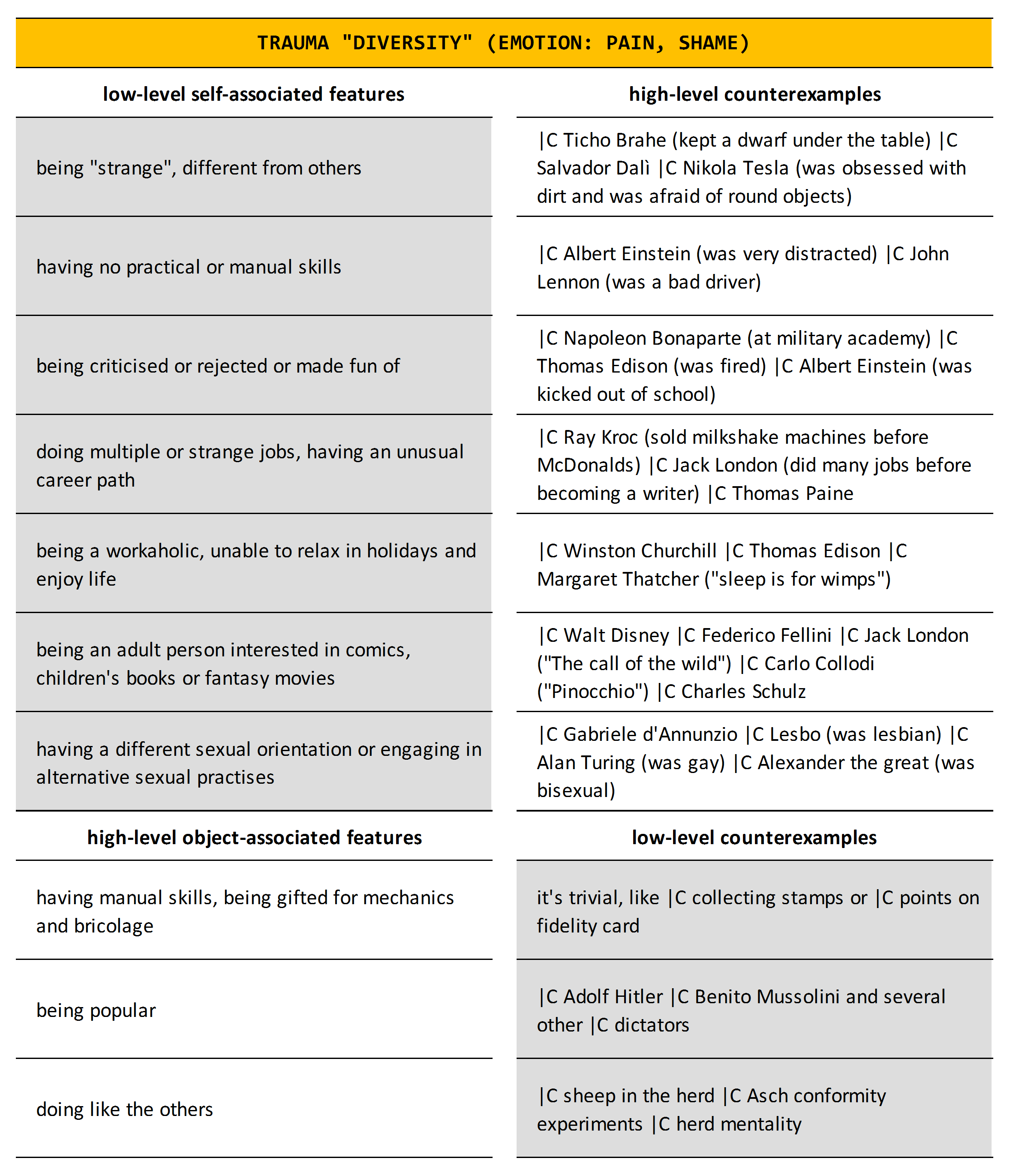}}}
\caption{Counterexamples (symbol ``$\mid$C'') for features linked to trauma ``DIVERSITY''. The information reported is correct to the best of the authour's knowledge, does not refer to any specific ``self'' or ``object'', is intended for scientific use only and does not intend to offend any person or entity. As a further precaution, only historical figures and fictional characters are used as examples.}
\label{diversity}
\end{center} \end{figure}

\begin{figure}[p] \begin{center} \hspace*{-0.25cm}
{\fboxrule=0.0mm\fboxsep=0mm\fbox{\includegraphics[width=17.50cm]{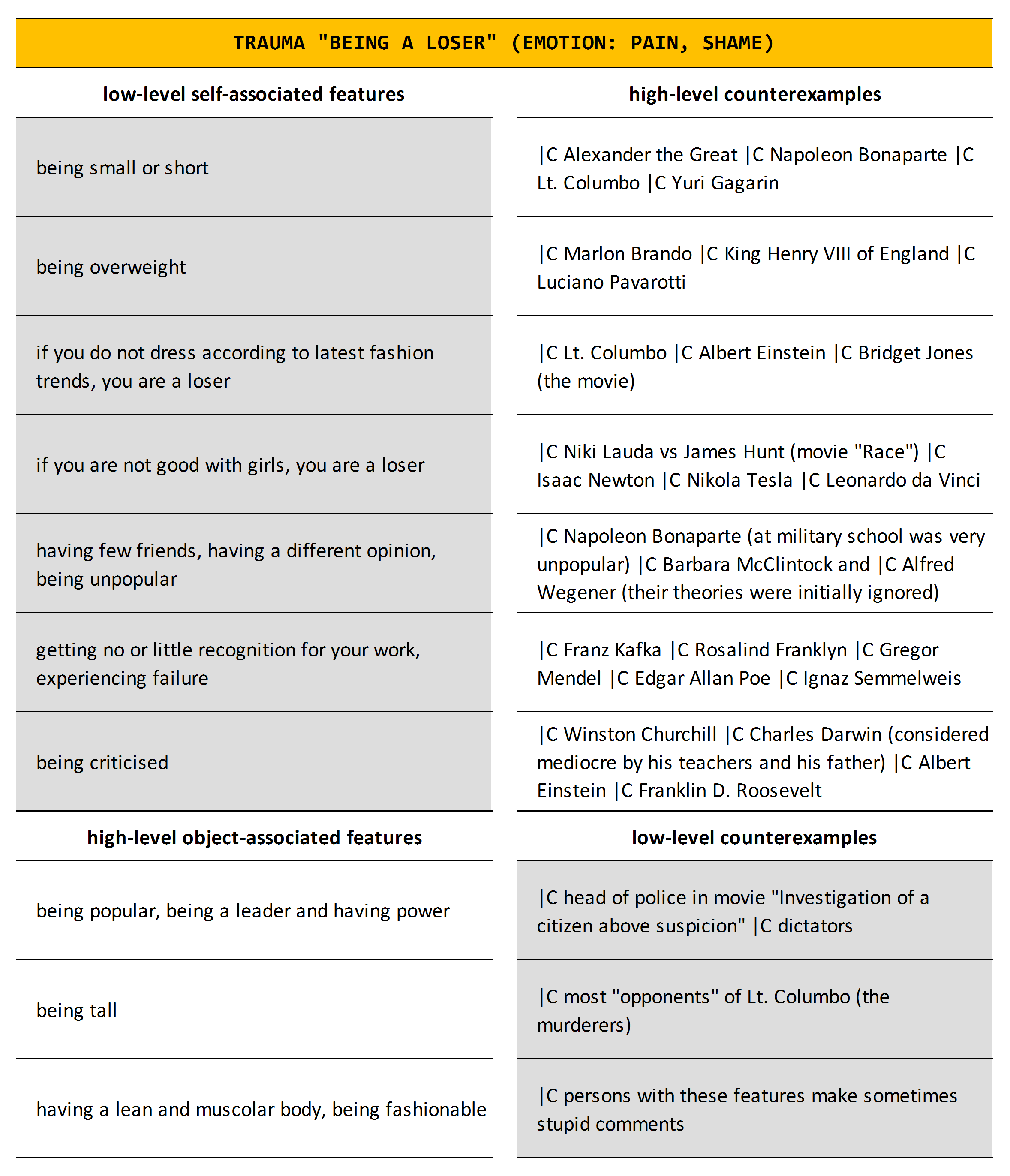}}}
\caption{Counterexamples (symbol ``$\mid$C'') for features linked to trauma ``BEING A LOSER''. The information reported is correct to the best of the authour's knowledge, does not refer to any specific ``self'' or ``object'', is intended for scientific use only and does not intend to offend any person or entity. As a further precaution, only historical figures and fictional characters are used as examples.}
\label{loser}
\end{center} \end{figure}

\clearpage

\section{Conclusions}

\ifhpar \colorbox{colhd}{xxxx} \\ \fi
The objective of this work was to propose a new psychotherapy for psychological traumas. The core idea is to make a list of self-associated low value ideas linked to each trauma and to provide for each idea a list of high value counterexamples, with the objective to raise the self value and solve the trauma. The psycotherapy proposed has not been tested on a clinical population yet, therefore statements on its effectiveness are premature. However, the conceptual basis is solid and, since traumas are hypothesised to be present in most psychological disorders, the potential gain may be substantial.

\section{Disclosure of interest}

The author is an employee of the European Research Council Executive Agency. The views expressed are purely those of the writer and may not in any circumstances be regarded as stating an official position of the European Commission.

\bibliographystyle{apalike}
\bibliography{ltheraprop} 

\end{document}